\documentclass[reprint,aps,showpacs,pra,nofootinbib]{revtex4-1}
  \usepackage{color}
  \usepackage{newlfont}
  \usepackage{graphicx}
  \usepackage{amssymb}
  \usepackage{amsmath}
  \usepackage{bm}
  \usepackage{verbatim}
  \usepackage[latin1]{inputenc}
  \usepackage{bbm}
  \usepackage{multirow}
  \usepackage[thinlines]{easytable}
 \usepackage[varg]{txfonts} 

\newcommand{\ket}[1]{\mbox{$ | #1 \rangle $}}
\newcommand{\bra}[1]{\mbox{$ \langle #1 | $}} 
\newcommand{\be}{\begin{equation}}
\newcommand{\ee}{\end{equation}}
\newcommand{\beq}{\begin{eqnarray}}
\newcommand{\eeq}{\end{eqnarray}}

\begin{document}

\title{Resilience of realism-based nonlocality to local disturbance}
\author{V. S. Gomes}
\author{R. M. Angelo}
\affiliation{Department of Physics, Federal University of Paran\'a, P.O. Box 19044, 81531-980 Curitiba, Paran\'a, Brazil}

\begin{abstract} 
Employing a procedure called monitoring---via a completely positive trace-preserving map that is able to interpolate between weak and projective measurements---we investigate the resilience of the recently proposed realism-based nonlocality to local and bilocal weak measurements. This analysis indicates realism-based nonlocality as the most ubiquitous and persistent form of quantumness within a wide class of quantum-correlation quantifiers. In particular, we show that the set of states possessing this type of quantumness forms a strict superset of symmetrically discordant states and, therefore, of discordant, entangled, steerable, and Bell-nonlocal states. Moreover, we find that, under monitoring, realism-based nonlocality is not susceptible to sudden death.
\keywords{Realism-based nonlocality \and Weak measurements \and Weak realism-based nonlocality}
\end{abstract}


\maketitle

\section{Introduction} 
By now it is safe to say that the notion of Bell nonlocality, as referring to a violation of Bell's hypothesis of local causality \cite{bell64}, is very well defined formally \cite{brunner14} and convincingly demonstrated experimentally \cite{hensen15,giustina15,shalm15,hensen16}, although some debate persists about the real meaning of such violations \cite{gisin12,tipler14}. It is also fair to state that these developments have been triggered by a firm notion of local realism, which was shared by many physicists of the beginning of the twentieth century and led Einstein, Podolsky, and Rosen (EPR) to suggest that quantum mechanics could not be the whole story about nature~\cite{EPR35}.

Recently, a notion of nonlocality has been put forward that profoundly differs from Bell nonlocality \cite{gomes18}. The concept was constructed by using, as primitive mechanism, a quantifier of the irreality degree \cite{bilobran15} 
\be \label{frakI}
\mathfrak{I}(A|\rho):=S(\Phi_A(\rho))-S(\rho)=S(\rho||\Phi_A(\rho))
\ee 
of an observable $A$ acting on $\cal{H_A}$ given a preparation $\rho$ on $\cal{H_A\otimes H_B}$, where $S(\rho)$ stands for the von Neumann entropy of $\rho$ and $S(\rho||\sigma)$ is the quantum relative entropy of $\rho$ and $\sigma$. Roughly, the irreality \eqref{frakI} gives an entropic distance between the state $\rho$ under scrutiny and a state
\be \label{Phi}
\Phi_A(\rho):=\sum_aA_a\,\rho\,A_a=\sum_ap_aA_a\otimes\rho_{\cal{B}|a},
\ee 
where $p_a=\text{Tr}(A_a\otimes\mathbbm{1}\,\rho)$ and $\rho_{\cal{B}|a}=\text{Tr}_{\cal{A}}(A_a\otimes\mathbbm{1}\rho)/p_a$. The state $\Phi_A(\rho)$ can be viewed as resulting from a projective measurement of the observable $A=\sum_aaA_a$ (where $A_aA_{a'}=\delta_{aa'}A_a$) in a scenario where the outcome is omitted. Since $A$ has been measured and thus became well definite, $\Phi_A(\rho)$ is a state of reality for $A$. Hence, $\mathfrak{I}(A|\rho)$ quantifies the violation of the hypothesis of realism, $\Phi_A(\rho)=\rho$. Notice, in particular, that the reduced state $\rho_{\cal{A}}=\text{Tr}_{\cal{B}}\Phi_A(\rho)=\sum_ap_aA_a$ has no quantum coherence $A$'s eigenbasis, that is, it is just a mixture of states $A_a$ with elements of reality $a$ for $A$. In addition, $\Phi_A\Phi_A(\rho)=\Phi_A(\rho)$, showing that further unrevealed measurements do not change a preestablished reality. Finally, since $\Phi_A(\rho)$ is a completely positive trace-preserving map, it follows from the monotonicity of the von Neumann entropy that $\mathfrak{I}(A|\rho)\geqslant 0$, with the equality holding for $\rho=\Phi_A(\rho)$. The work \cite{gomes18} then focused on the difference 
\beq \label{eta}
\eta_{AB}(\rho)&:=&\mathfrak{I}(A|\rho)-\mathfrak{I}(A|\Phi_B(\rho)),
\eeq 
which measures alterations in the irreality of $A$ induced by unrevealed projective measurements of $B$ conducted in a remote site $\cal{B}$. This object has been introduced in Ref. \cite{bilobran15} as a measure of the realism-based nonlocality of a context defined by $\{A,B,\rho\}$, as it quantifies violations of the hypothesis of local realism, $\mathfrak{I}(A|\rho)=\mathfrak{I}(A|\Phi_B(\rho))$, for a particular context. At this point we can readily recognize the conceptual difference between Bell nonlocality and realism-based nonlocality: the former accuses violations of the local causality hypothesis, whereas the latter refers to a violation of local realism in the terms delineated above. Using the definition \eqref{frakI}, we also note that
\be\label{etaS} 
\eta_{AB}(\rho)=S(\Phi_A(\rho))+S(\Phi_B(\rho))-S(\Phi_A\Phi_B(\rho))-S(\rho),
\ee 
which reveals the presence of the symmetry $A\rightleftarrows B$. This means that $\eta_{AB}(\rho)$ also quantifies changes in $B$'s irreality induced by unrevealed measurements of $A$ in a remote site. Finally, the realism-based nonlocality $N(\rho)$ of a preparation $\rho$ was introduced as the maximum difference $\eta_{AB}(\rho)$ over all possible pair of observables $\{A,B\}$ measured in distant sites:
\be \label{N}
N(\rho):=\max_{A,B}\,\eta_{AB}(\rho).
\ee 
As pointed out in Refs. \cite{gomes18,bilobran15}, $\eta_{AB}$ is a nonnegative quantity that vanishes only for fully uncorrelated states ($\rho=\rho_{\cal{A}}\otimes\rho_{\cal{B}}$) or for states $\rho=\Phi_{A(B)}(\rho)$ with reality for $A$ ($B$). In addition, $N$ vanishes for $\rho=\rho_{\cal{A}}\otimes\rho_{\cal{B}}$ and reduces to the entanglement entropy for pure states, thus being free of the anomaly that affect some Bell-nonlocality measures \cite{acin02,methot07,vidick11,camaleT27}.

Unlike the Bell nonlocality framework, for which much progress has already been reached (see, e.g., Ref. \cite{brunner14} for formal characterization and related concepts, applications, experimental aspects, and multipartite extensions, Ref. \cite{vicente14} for a resource theory and quantifiers, and Ref. \cite{buscemi12} for the connection with entanglement), the research on realism-based nonlocality is rather incipient. Among several open questions concerning this quantity, one may cite the search for a resource theory, operational meaning, extension to multipartite scenarios, and applications in information-theoretic or thermodynamic tasks. Also, it is not known yet how realism-based nonlocality behaves under local operations. Such a study is crucial to assess whether realism-based nonlocality can be viewed as a genuine measure of nonlocal quantum correlations. In particular, one may wonder whether realism-based nonlocality is susceptible to sudden death, a phenomenon that has already been predicted for entanglement, EPR steering, and Bell nonlocality (see., e.g., Ref. \cite{costa16}).

With the aim of initiating this program, in this work we investigate the robustness of realism-based nonlocality under some classes of local operations. We ask how $N(\rho)$ changes when monitorings \cite{dieguez18} (that is, weak measurements \cite{aharonov88,aharonov14,oreshkov05} without postselection) are locally conducted in one or both sites. It is shown via rigorous results and case studies that there can be no abrupt variations of realism-based nonlocality under monitorings, meaning that these nonlocal aspects are only smoothly suppressed under weak disturbances. In this sense, realism-based nonlocality can be diagnosed as a quantumness aspect that is more resilient than Bell nonlocality.

\section{Theoretical preliminaries}
We start by indicating the model of weak measurements we shall employ throughout this work. In Ref. \cite{dieguez18} a procedure was introduced---the so-called monitoring---which is mathematically described by the completely positive trace-preserving map
\be\label{Me}
M_O^{\epsilon}(\rho):=(1-\epsilon)\,\rho+\epsilon\,\Phi_O(\rho),
\ee 
for a real parameter $\epsilon\in(0,1)$, $\rho$ on $\cal{H_A\otimes H_B}$, and a generic observable $O$ acting on either $\cal{H_A}$ or $\cal{H_B}$. This map continuously interpolates between a regime of no measurement at all ($M_O^{\epsilon\to 0}=\mathbbm{1}$) to the one of unrevealed projective measurements ($M_O^{\epsilon\to 1}=\Phi_O$). Between these extrema, that is, for $0<\epsilon<1$, the map is then said to correspond to an unrevealed weak measurement. The adjective ``unrevealed'' emphasizes that the outcome of the measurement is not accessed by the experimentalist, so that no collapse or information gain occur. In this sense, the map $M_O^{\epsilon}$ differs from the standard weak-measurement approach that is defined via conjugation of weak interactions with post selections. Despite its relevance, the term ``unrevealed'' will be henceforth omitted. As pointed out in Ref. \cite{dieguez18}, the monitoring of $O$ can also be thought of as deriving from a von Neumann premeasurement through which this quantity gets correlated with an ancillary system that is posteriorly discarded. By direct application of the definitions \eqref{Phi} and \eqref{Me} one shows that
\be \label{MPhi}
M_O^{\epsilon}\Phi_O=\Phi_OM_O^{\epsilon}=\Phi_O.
\ee 
Successive applications of the same monitoring yields 
\be \label{Men}
[M_O^{\epsilon}]^n(\rho)=(1-\epsilon)^n\rho+\big[1-(1-\epsilon)^n\big]\Phi_O(\rho),
\ee
which implies that $[M_O^{\epsilon}]^{n\to\infty}=\Phi_O$. This shows that a sequence of infinitely many monitorings is equivalent to a projective unread measurement. Most importantly, it can be shown  \cite{bilobran15,dieguez18} that $S(\Phi_O(\rho))\geqslant S([M_O^{\epsilon}]^n(\rho))$, for any finite integer $n$, with the equality holding for $\rho=\Phi_O(\rho)$, and 
\be \label{monotfrakI}
\mathfrak{I}(A|\rho)\geqslant \mathfrak{I}(A|M_O^{\epsilon}(\rho)), 
\ee
for $A$ on $\cal{H_A}$. This relation is a statement of the monotonicity of the irreality under local monitoring. It means that upon a weak measurement of arbitrary intensity, in whatever location, the reality of an observable never decreases. 

The dual of irreality has recently been discussed in Ref. \cite{dieguez18}, where the authors introduced the quantifier 
\be \label{DR}
\Delta \mathfrak{R}_A^{\epsilon}(\rho):=S(M_A^{\epsilon}(\rho))-S(\rho)
\ee 
for the weak increase in the reality of a monitored observable $A$ given the preparation $\rho$. Clearly, there is no reality increase for $\epsilon\to 0$ and maximum increase of $\mathfrak{I}(A|\rho)$ for $\epsilon\to 1$. Recently, using state tomography Mancino {\it et al.} measured $\Delta\mathfrak{R}$ for the polarization of a photon as a function of the intensity $\epsilon$ of the monitoring in a photonics experiment~\cite{mancino18}.

In connection with the context realism-based nonlocality \eqref{etaS}, and for future convenience, we prove a useful result for the following quantity:
\be \label{delta}
\delta_{AB}^{\epsilon\epsilon'}(\rho):=S(M_A^{\epsilon}(\rho))+S(M_B^{\epsilon'}(\rho))-S(M_A^{\epsilon}M_B^{\epsilon'}(\rho))-S(\rho),
\ee 
where $M_{A(B)}^{\epsilon(\epsilon')}$ is a local monitoring of the observable $A(B)$ on $\cal{H_{A(B)}}$ with intensity $\epsilon (\epsilon')$ and $\{\epsilon,\epsilon'\}\in(0,1)$. Notice that $\delta_{AB}^{11}(\rho)=\eta_{AB}(\rho)$ and $\delta_{AB}^{0\epsilon'}(\rho)=\delta_{AB}^{\epsilon 0}(\rho)=0$, where it should be understood that these terms have been calculated in the limit as $\epsilon(\epsilon')$ goes to 0 or 1.

\vskip2mm\noindent {\it Theorem 1 \label{T1}}. For any density operator $\rho$ on $\cal{H_A\otimes H_B}$, observables $A$ on $\cal{H_A}$ and $B$ on $\cal{H_B}$, and $\{\epsilon,\epsilon'\}\in (0,1)$, it holds that $\delta_{AB}^{\epsilon\epsilon'}(\rho)\geqslant 0$. Equality applies for $\rho=\rho_{\cal A}\otimes \rho_{\cal B}$ or $\rho=\Phi_{A(B)}(\rho)$.

\vskip2mm\noindent {\it Proof.} Consider $\sigma=U\sigma_0U^{\dag}$ on $\cal{H_A\otimes H_B\otimes H_X\otimes H_Y}$, with the unitary transformation $U=U_{\cal{AX}}^{\epsilon}\otimes U_{\cal{BY}}^{\epsilon'}$ and the initial state $\sigma_0=\rho\otimes\ket{x}\bra{x}\otimes\ket{y}\bra{y}$. The strong subadditivity of the entropy, $S(\sigma_{abc})+S(\sigma_c)\leqslant S(\sigma_{ac})+S(\sigma_{bc})$ \cite{nielsen00} and the identifications $a=\cal{X}$, $b=\cal{Y}$, and $c=\cal{AB}$, allow us to write
\be 
S(\sigma)+S(\sigma_{\cal{AB}})\leqslant S(\sigma_{\cal{ABX}})+S(\sigma_{\cal{ABY}}).
\ee 
Via the unitary invariance and the additivity of the entropy, one has $S(\sigma)=S(\rho)$. Using $M_A^{\epsilon}(\rho)=\text{Tr}_{\cal{X}}[U_{\cal{AX}}^{\epsilon}\rho\otimes\ket{x}\bra{x}U_{\cal{AX}}^{\epsilon\dag}]$ and the analog relation for $M_B^{\epsilon'}(\rho)$, which are expressions of the Stinespring theorem \cite{dieguez18,nielsen00}, we derive the reduced states 
$\sigma_{\cal{ABX}}=U_{\cal{AX}}^{\epsilon}M_B^{\epsilon'}(\rho)\otimes\ket{x}\bra{x}U_{\cal{AX}}^{\epsilon\dag}$, $\sigma_{\cal{ABY}}=U_{\cal{BY}}^{\epsilon'}M_A^{\epsilon}(\rho)\otimes\ket{y}\bra{y}U_{\cal{BY}}^{\epsilon'\dag}$, and $\sigma_{\cal{AB}}=M_A^{\epsilon}M_B^{\epsilon'}(\rho)$. Using unitary invariance and additivity once more we find 
\be 
S(\rho)+S(M_A^{\epsilon}M_B^{\epsilon'}(\rho))\leqslant S(M_B^{\epsilon'}(\rho))+S(M_A^{\epsilon}(\rho)),
\ee 
which proves the first part of the theorem. From the additivity of the entropy, the property \eqref{MPhi}, and the definition \eqref{delta} it follows that $\delta_{AB}^{\epsilon\epsilon'}(\rho_{\cal A}\otimes\rho_{\cal B})=\delta_{AB}^{\epsilon\epsilon'}(\Phi_{A(B)}(\rho))=0$, which completes the proof\footnote{The equality also holds for $\epsilon\to 0$ or $\epsilon'\to 0$, but these limits are not included in the statement of the theorem.}.\hfill{$\blacksquare$}

Before proceeding, it is worth noticing that, from the definition \eqref{delta} and the property \eqref{MPhi}, we can directly demonstrate the following identities:
\be \label{difdelta}
\delta_{AB}^{\epsilon\epsilon'}(\rho)=\delta_{AB}^{\epsilon 1}(\rho)-\delta_{AB}^{\epsilon 1}(M_B^{\epsilon'}(\rho))=\delta_{AB}^{1\epsilon'}(\rho)-\delta_{AB}^{1\epsilon'}(M_A^{\epsilon}(\rho)).
\ee 
These relations reveal an interesting point. One might be tempted at a first sight to interpret $\delta_{AB}^{\epsilon\epsilon'}(\rho)$ as mathematical extension of the context realism-based nonlocality \eqref{etaS} to the regime of monitorings, after all, $\delta_{AB}^{11}(\rho)=\eta_{AB}(\rho)$. However, one has that $\delta_{AB}^{0\epsilon'}(\rho)=\delta_{AB}^{\epsilon 0}(\rho)=\delta_{AB}^{00}(\rho)=0$ even when $\eta_{AB}(\rho)>0$; that is, this interpretation would cause a clear incompatibility among the predictions given by the quantifiers $\eta_{AB}$ and $\delta_{AB}^{\epsilon\epsilon'}$. It follows, therefore, that $\delta_{AB}^{\epsilon\epsilon'}(\rho)$ is to be interpreted in a rather different way. Let us specialize the analysis of Eq. \eqref{difdelta} to the quantity $\delta_{AB}^{1\epsilon}(\rho)=\eta_{AB}(\rho)-\eta_{AB}(M_B^{\epsilon}(\rho))$ for a while. Since $\eta_{AB}(\rho)$ and $\eta_{AB}(M_B^{\epsilon}(\rho))$ are quantifiers of the amount of realism-based nonlocality for the contexts $\{A,B,\rho\}$ and $\{A,B,M_B^{\epsilon}(\rho)\}$, respectively, then $\delta_{AB}^{1\epsilon}(\rho)$ can be interpreted as the amount of realism-based nonlocality that is suppressed when $\rho$ is replaced with its monitored counterpart $M_B^{\epsilon}(\rho)$. In other words, $\delta_{AB}^{1\epsilon}(\rho)$ turns out to be a measure of the amount of realism-based nonlocality that is destroyed upon local monitoring of the observable $B$ with intensity $\epsilon$. Accordingly, we see that no realism-based nonlocality is destroyed for $\epsilon\to 0$ whereas all realism-based nonlocality is destroyed for $\epsilon\to 1$. This can also be verified through the following formulation. First, using precedent formulas notice that $\delta_{AB}^{1\epsilon}(M_B^{\epsilon}(\rho))=\eta_{AB}(M_B^{\epsilon}(\rho))-\eta_{AB}([M_B^{\epsilon}]^2(\rho))$ and, therefore, $\delta_{AB}^{1\epsilon}(\rho)+\delta_{AB}^{1\epsilon}(M_B^{\epsilon}(\rho))=\eta_{AB}(\rho)-\eta_{AB}([M_B^{\epsilon}]^2(\rho))$, which, via Eq. \eqref{Men}, can be identified to $\delta_{AB}^{1\epsilon'}(\rho)$ with $\epsilon'=1-(1-\epsilon)^2$. By summing terms for higher-order monitorings we find
\be 
\sum_{k=0}^{n}\delta_{AB}^{1\epsilon}([M_B^{\epsilon}]^k(\rho))=\delta_{AB}^{1\epsilon_n}(\rho) \stackrel{(n\to\infty)}{\longrightarrow} \eta_{AB}(\rho),
\ee 
where $\epsilon_n=1-(1-\epsilon)^{n+1}$. This shows that adding the amount $\delta_{AB}^{1\epsilon}$ of realism-based nonlocality that is suppressed upon $\epsilon$-intensity monitorings $[M_B^{\epsilon}]^k$ of infinitely many orders $k$ is equal to the total amount $\eta_{AB}(\rho)$ associated with the state $\rho$.

\section{Resilience to local weak measurement}
\subsection{Realism-based nonlocality suppression via local monitoring}

We now introduce and analyze a quantifier that will prove informative with respect to the resilience of realism-based nonlocality under local monitoring. Inspired by the previous discussion, we take
\be \label{D_B}
\Delta_{\cal B}^{\epsilon}(\rho):=\max_{A,B,B'}\Big[\eta_{AB}(\rho)-\eta_{AB}(M_{B'}^{\epsilon}(\rho))\Big] 
\ee 
as a quantifier of the optimized suppression implied to the context realism-based nonlocality when the state $\rho$ on $\cal{H_A\otimes H_B}$ is submitted to a local monitoring in the site $\cal{B}$. An analog quantity $\Delta_{\cal A}^{\epsilon}$ can be constructed for monitorings in the site $\cal{A}$, and it should be clear, by construction, that it is not necessary that $\Delta_{\cal A}^{\epsilon}=\Delta_{\cal B}^{\epsilon}$ for generic states. Notice that $\Delta_{\cal{B}}^0(\rho)=0$, meaning that no realism-based nonlocality is destroyed when the system suffers no monitoring. The quantity \eqref{D_B} can be simplified. With Eq. \eqref{eta}, we write
\be \label{aux}
\Delta_{\cal B}^{\epsilon}(\rho)=\max_{A,B,B'}\Big[\mathfrak{I}(A|\rho)-\mathfrak{I}(A|M_{B'}^{\epsilon}(\rho))-\chi\Big],
\ee
where $\chi\equiv\mathfrak{I}(A|\Phi_B(\rho))-\mathfrak{I}(A|\Phi_BM_{B'}^{\epsilon}(\rho))$. The relation \eqref{monotfrakI} then implies that $\chi\geqslant 0$. Hence, the maximization in Eq. \eqref{aux} with respect to $B'$ is attained for $\chi=0$. According to Eq. \eqref{MPhi}, this can be ensured for a generic $\rho$ via the choice $B'=B$. Using the relation $\delta_{AB}^{1\epsilon}(\rho)=\eta_{AB}(\rho)-\eta_{AB}(M_B^{\epsilon}(\rho))$ one obtains
\be \label{De}
\Delta_{\cal B}^{\epsilon}(\rho)=\max_{A,B}\Big[\eta_{AB}(\rho)-\eta_{AB}(M_B^{\epsilon}(\rho))\Big]=\max_{AB}\delta_{AB}^{1\epsilon}(\rho),
\ee 
which will be the figure of merit considered from now on. By virtue of Theorem 1 we have $\Delta_{\cal B}^{\epsilon}(\rho)\geqslant 0$ for any $\epsilon\in(0,1)$. Because $\eta_{AB}$ is nonnegative, it is also clear that $\Delta_{\cal B}^{\epsilon}(\rho)\leqslant N(\rho)$. This has to be so, since the amount of realism-based nonlocality destroyed by the monitoring can never be grater than the available realism-based nonlocality. Summing up, we have shown that the bounds
\be \label{bounds}
0\leqslant \Delta_{\cal B}^{\epsilon}(\rho)\leqslant N(\rho)
\ee 
hold for $\epsilon\in(0,1)$ and all $\rho$ on $\cal{H_A\otimes H_B}$, with the equalities simultaneously applying for $\rho=\rho_{\cal A}\otimes\rho_{\cal B}$. In what follows we derive tighter bounds. To this end, we will employ the relation 
\be \label{DS<}
S(M_B^{\epsilon}(\rho))-S(\rho)\leqslant \Gamma_B^{\epsilon}(\rho),
\ee
with $\Gamma_B^{\epsilon}(\rho)\equiv \epsilon\tau_B(\rho)\ln{(d-1)}+H(\epsilon\tau_B(\rho))$, which was proved in Ref. \cite{dieguez18}. Here, $H(p)=-p\ln{p}-(1-p)\ln{(1-p)}$ is the Shannon entropy, $\rho$ is a density operator on a space $\cal{H}$ with dimension is $d$, and $\tau_B(\rho)=T(\Phi_B(\rho),\rho)$ is the trace norm, where $T(\rho,\sigma)=\tfrac{1}{2}\text{Tr}||\rho-\sigma||_1\in[0,1]$ and $||\varrho||_1=(\varrho^{\dag}\varrho)^{1/2}$. Reference \cite{dieguez18} also points out that $\Gamma_B^{\epsilon}(\rho)<d[\epsilon\tau_B(\rho)/e]^{1/2}$, $e$ being the Euler number, thus showing that the upper bound $\Gamma_B^{\epsilon}(\rho)$ is never greater than $O(\epsilon^{\text{\tiny $1/2$}})$. The equality in the relation \eqref{DS<} applies when $\rho=\Phi_B(\rho)$, in which case both sides vanish. On the other hand, the concavity of the entropy gives 
\be \label{DS>}
S(M_B^{\epsilon}(\rho))-S(\rho)\geqslant\epsilon\,\mathfrak{I}(B|\rho),
\ee 
with the equality holding for $\rho=\Phi_B(\rho)$. Now, taking the formula \eqref{delta} to compute $\delta_{AB}^{1\epsilon}$, we can apply the above inequalities (with pertinent adaptations) to the terms $S(M_B^{\epsilon}(\rho))-S(\rho)$ and $S(M_B^{\epsilon}\Phi_A(\rho))-S(\Phi_A(\rho))$, respectively, so as to obtain an upper bound to $\delta_{AB}^{1\epsilon}$ and to $\Delta_{\cal B}^{\epsilon}$. On the other hand, applying the inequality \eqref{DS>} to the first term and \eqref{DS<} to the second, we obtain lower bounds.  Thus, via Eq. \eqref{De} we arrive at 
\begin{subequations}\label{result1}
\beq 
\Delta_{\cal B}^{\epsilon}(\rho)&\leqslant& \Gamma_{\bar{B}}^{\epsilon}(\rho)-\epsilon\,\mathfrak{I}(\bar{B}|\Phi_{\bar{A}}(\rho))\equiv \mathrm{UB_1}, \\ \Delta_{\cal B}^{\epsilon}(\rho)&\geqslant&\epsilon\,\mathfrak{I}(\bar{B}|\rho)-\Gamma_{\bar{B}}^{\epsilon}(\Phi_{\bar{A}}(\rho))\equiv \mathrm{LB}_1,
\eeq 
\end{subequations} 
where $\{\bar{A},\bar{B}\}$ are the observables that maximize $\delta_{AB}^{1\epsilon}(\rho)$ in Eq. \eqref{De}. Further bounds can be derived as follows. By virtue of the monotonicity of the entropy under completely positive trace-preserving maps, one has $S(\Phi_B(\rho))\geqslant S(M_B^{\epsilon}(\rho))\geqslant S(\rho)$, with the equality holding for $\rho=\Phi_B(\rho)$. Along with the definition \eqref{frakI}, this allows us to write 
\be \label{SI<}
S(M_B^{\epsilon}(\rho))-S(\rho)\leqslant \mathfrak{I}(B|\rho).
\ee 
Applying this inequality and the one \eqref{DS>} to the term $S(M_B^{\epsilon}\Phi_A(\rho))-S(\Phi_A(\rho))$ we find an upper bound $\mathrm{UB}_2$ to $\Delta_{\cal B}^{\epsilon}$. On the other hand, by direct use of the relation \eqref{DS>} and adapting the inequality \eqref{SI<} to $S(M_B^{\epsilon}\Phi_A(\rho))-S(\Phi_A(\rho))$, we arrive at a lower bound $\mathrm{LB}_2$. These results are written as
\begin{subequations} \label{result2}
\beq 
\Delta_{\cal B}^{\epsilon}(\rho)&\leqslant& \epsilon\,\eta_{\bar{A}\bar{B}}(\rho)+(1-\epsilon)\mathfrak{I}(\bar{B}|\rho)\equiv \mathrm{UB}_2, \\
\Delta_{\cal B}^{\epsilon}(\rho)&\geqslant& \epsilon\,\eta_{\bar{A}\bar{B}}(\rho)-(1+\epsilon)\mathfrak{I}(\bar{B}|\Phi_{\bar{A}}(\rho))\equiv \mathrm{LB}_2.
\eeq  
\end{subequations}
Clearly, while $\Delta_{\cal B}^{\epsilon\to 0}(\rho)=0$ these bounds do not necessarily vanish with $\epsilon$, so they cannot always be tight. Yet, as will be numerically shown later, each one of the bounds derived above may reveal its particular superiority in specific instances. 

The set of bounds defined by the relations \eqref{bounds}, \eqref{result1}, and \eqref{result2} constitute the fundamental result of this work. $\mathrm{LB_1}$, in particular, guarantees that no matter how tiny the monitoring is there will always be some suppression of realism-based nonlocality, which is at least of the order $O(\epsilon)$. Therefore, it is certain that realism-based nonlocality will not increase under monitoring. On the other hand, the suppression will never equal the total realism-based nonlocality of the state and can be limited to the order $O(\epsilon)$, as indicated by $\mathrm{UB}_1$. As a consequence, even when the amount of realism-based nonlocality is small for a given state, the local monitoring will never destroy all the available realism-based nonlocality, which then implies no sudden death whatsoever. 

As far as pure states $\varsigma=\ket{\psi}\bra{\psi}$ are concerned, one can show that the suppression \eqref{De} of realism-based nonlocality under local monitoring is bounded as
\be \label{result3}
\epsilon\,E(\varsigma)\leqslant \Delta_{\cal B}^{\epsilon}(\varsigma)\leqslant E(\varsigma),
\ee 
where $E(\varsigma)=S(\mathrm{Tr}_{\cal{A(B)}}\varsigma)$ is the entanglement entropy of $\ket{\psi}$. The proof goes as follows. Since $N(\varsigma)=E(\varsigma)$ for any pure state $\varsigma=\ket{\psi}\bra{\psi}$ (see Ref. \cite{gomes18}), the second inequality above trivially follows from the bounds \eqref{bounds}. To prove the first inequality, we use Eq. \eqref{De}, the definition of irreality \eqref{frakI},  the fact that $S(\varsigma)=0$, and the monotonicity relation $S(\Phi_A\Phi_B(\varsigma))\geqslant S(\Phi_AM_B^{\epsilon}(\varsigma))$. All this allows us to write
\be
\Delta_{\cal B}^{\epsilon}(\varsigma)\geqslant \max_{A,B}\left\{S(M_B^{\epsilon}(\varsigma))-\Big[S(\Phi_A\Phi_B(\varsigma))-S(\Phi_A(\varsigma))\Big]\right\},
\ee
where the equality holds for $\Phi_B(\varsigma)=M_B^{\epsilon}(\varsigma)$. Since the term in square brackets is nonnegative, the maximization will occur for Schmidt operators $A_S=\sum_ia_i\ket{i}\bra{i}$ and $B_S=\sum_ib_i\ket{i}\bra{i}$ (the ones that define the basis $\{\ket{i}\ket{i}\}$ which leads to the Schmidt decomposition $\ket{\psi}=\sum_i\sqrt{\lambda_i}\ket{i}\ket{i}$), for in this case it follows that $S(\Phi_{A_S}\Phi_{B_S}(\varsigma))-S(\Phi_{A_S}(\varsigma))=0$. Then, using the definition \eqref{Me} and the concavity of entropy we find
\be
\Delta_{\cal B}^{\epsilon}(\varsigma)\geqslant S(M_{B_S}^{\epsilon}(\varsigma))\geqslant \epsilon\,S(\Phi_{B_S}(\varsigma)).
\ee 
The observation that $S(\Phi_{B_S}(\varsigma))=S(\text{Tr}_{\cal{A(B)}}\varsigma)=E(\varsigma)$ completes the proof. It is worth noticing that by employing the Schmidt operators we can arrive at the result \eqref{result3} departing from the bounds \eqref{result2}, which certifies the consistence of the approach. The bounds \eqref{result3} show that local monitorings can never destroy more realism-based nonlocality (entanglement, for pure states) than the amount available in the preparation, but will destroy at least an amount $\epsilon E(\varsigma)$. Thus, just as in the case of mixed states, increase and sudden death of realism-based nonlocality are both prevented.

\subsubsection{Example}
\label{example}
Let us consider the two-parameter state of two qubits
\be \label{rhoab}
\rho^{\alpha\beta}=(1-\beta)\tfrac{\mathbbm{1}\otimes\mathbbm{1}}{4}+\beta\ket{\psi_{\alpha}}\bra{\psi_{\alpha}}, \qquad
\ee 
where $\ket{\psi_{\alpha}}=\sqrt{\alpha}\ket{01}-\sqrt{1-\alpha}\ket{10}$ and $\{\alpha,\beta\}\in [0,1]$. 

\begin{figure}[htb]
\centerline{\includegraphics[scale=0.85]{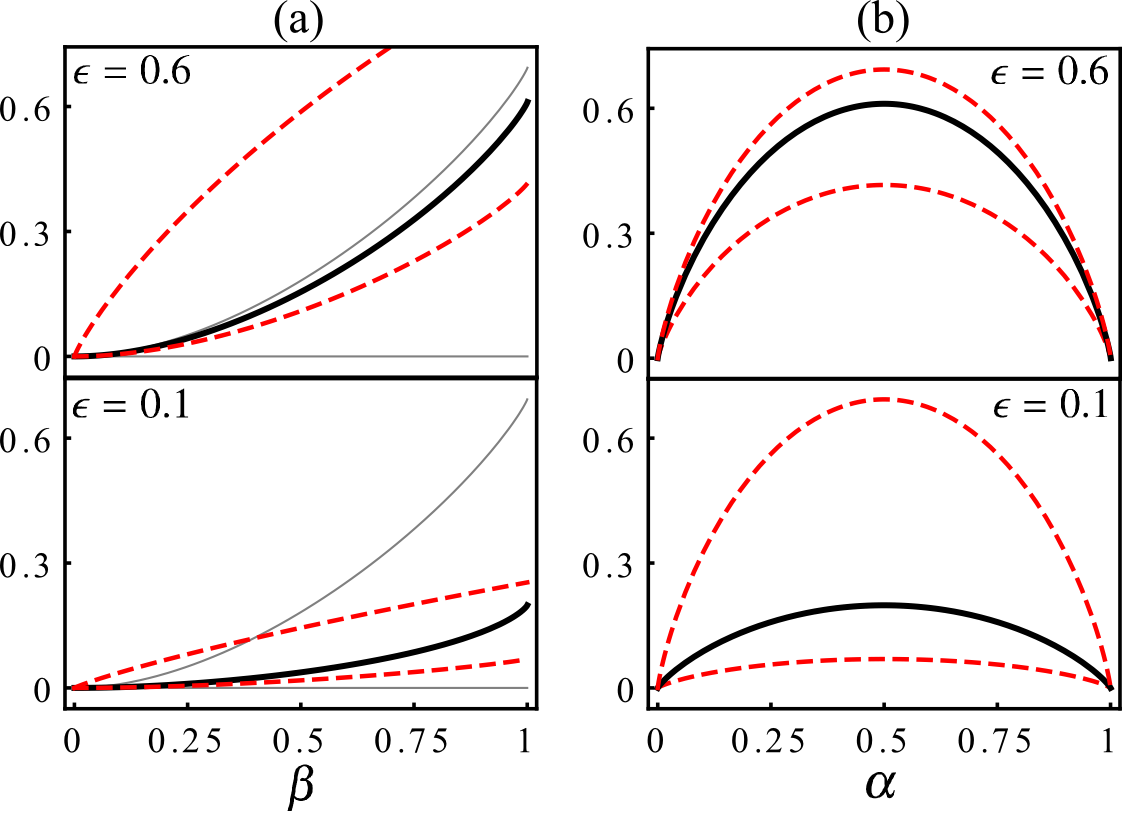}}
\caption{Amount $\Delta_{\cal B}^{\epsilon}$ (thick black lines) of realism-based nonlocality that is removed from the state $\rho^{\alpha\beta}$ [Eq. \eqref{rhoab}] for local monitorings of intensities $\epsilon=0.1$ (lower panels) and $\epsilon=0.6$ (upper panels). (a) Illustration of the results \eqref{bounds}, \eqref{result1}, and \eqref{result2}. The bounds given by the relations \eqref{bounds} are plotted in thin gray lines, whereas the ones defined by \eqref{result1}, $\mathrm{LB_1}$ and $\mathrm{UB_1}$, appear in dashed red lines. For the state $\rho^{\text{\tiny $\tfrac{1}{2}$}\beta}$ under concern, we found that $\mathrm{LB_2=LB_1}$ and $\mathrm{UB_2}=N(\rho)$, so that the bounds given by the result \eqref{result2} provide no more information than the previous ones. (b) Illustration of the bounds \eqref{result3}, with the pure state $\rho^{\alpha 1}=\ket{\psi_{\alpha}}\bra{\psi_{\alpha}}$, where $\ket{\psi_{\alpha}}=\sqrt{\alpha}\ket{01}-\sqrt{1-\alpha}\ket{10}$. In each graph, the upper (lower) dashed red line denotes the upper (lower) bound $E(\rho^{\alpha 1})$ [$\epsilon E(\rho^{\alpha 1})$] indicated in the result \eqref{result3}.}
\label{fig1} 
\end{figure}

We first focus on the state $\rho^{\text{\tiny $\tfrac{1}{2}$}\beta}$, which corresponds to a Werner state with $\ket{\psi_{\text{\tiny $\frac{1}{2}$}}}$ being the singlet state. To compute the realism-based nonlocality suppression through the formulas \eqref{De} and \eqref{delta} we consider the generic observable $A=\sum_{a=\pm}aA_a$ with projectors $A_{\pm}=\ket{\pm}\bra{\pm}$ such that $\ket{+} =\cos{(\tfrac{\theta_a}{2})}\ket{0} +e^{i\phi_a}\sin{(\tfrac{\theta_a}{2})}\ket{1}$ and $\ket{-}=-\sin{(\tfrac{\theta_a}{2})}\ket{0} +e^{i\phi_a}\cos{(\tfrac{\theta_a}{2})}\ket{1}$, and a similar parametrization for $B$ in terms of $\{\phi_b,\theta_b\}$ and $\{\ket{0},\ket{1}\}\in\cal{H_B}$. Analytical calculations show that the maximum of $\delta_{AB}^{1\epsilon}$ in \eqref{De} is attained for $\theta_{a,b}=0$, which implies that $A=\sigma_z$ and $B=\sigma_z$. With that, we find 
\be \label{NeBrho12beta}
\Delta_{\cal B}^{\epsilon}(\rho^{\text{\tiny $\tfrac{1}{2}$}\beta})=\tfrac{1}{4}\sum_{i=0}^1\sum_{j=0}^1(-1)^{j}\lambda_{ij}^{\epsilon}\ln{\lambda_{ij}^{\epsilon}}.
\ee 
where $\lambda_{ij}^{\epsilon}=1+\beta\big[4i-1+2j\epsilon(1-2i)\big]$.
This function is plotted in Fig.~\ref{fig1}(a), where we can see that it indeed has the expected behavior: it is always less than realism-based nonlocality, increases with $\epsilon$, and is correctly positioned in between the bounds \eqref{bounds}, \eqref{result1}, and \eqref{result2}. For the state under scrutiny, we have also numerically verified that $\Delta_{\cal B}^{\epsilon}=\Delta_{\cal A}^{\epsilon}$.

We now consider the pure state $\rho^{\alpha 1}$. Instead of pursuing the analytical maximization demanded by the formula \eqref{De}, which is harder to perform in this case, we proceed by similarity. It has been shown in Ref. \cite{bilobran15} that 
\be \label{minimum}
\min_{A,B}\,\delta_{AB}^{11}(\varsigma)=\eta_{A_S\tilde{B}_S}^{11}(\varsigma)=0
\ee 
for any pure state $\varsigma=\ket{\psi}\bra{\psi}$, the Schmidt operator $A_S$, and the maximally incompatible operator $\tilde{B}_S$ (that is, the one for which the commutator $[\tilde{B}_S,B_S]$ is maximum). On the other hand, from Ref. \cite{gomes18} one has that 
\be \label{maximum}
\max_{A,B}\,\delta_{AB}^{11}(\varsigma)=\eta_{A_SB_S}^{11}(\varsigma)=E(\varsigma),
\ee 
which occurs for Schmidt operators $A=A_S$ and $B=B_S$. By taking $A=\sigma_z$ and $B=\sigma_x$ we have been able to analytically check that $\eta_{\sigma_z\sigma_x}^{1\epsilon}(\rho^{\alpha 1})=0$ for $\epsilon\in(0,1)$, just as in Eq. \eqref{minimum}. Following Ref. \cite{gomes18}, we then take $A=\sigma_z$ and $B=\sigma_z$, in order to attain higher values for $\Delta_{\cal B}^{\epsilon}$, and then conjecture, with basis in comparisons made with many other choices for $A$ and $B$, that $\Delta_{\cal B}^{\epsilon}(\rho^{\alpha 1})=\eta_{\sigma_z\sigma_z}^{1\epsilon}(\rho^{\alpha 1})$. This yields 
\be \label{DeBpure}
\Delta_{\cal B}^{\epsilon}(\rho^{\alpha 1})=-\ln{\sqrt{\Lambda_{\epsilon}}}-\sqrt{1-4\Lambda_{\epsilon}}\,\,\mathrm{arctanh}{\sqrt{1-4\Lambda_{\epsilon}}},
\ee 
where $\Lambda_{\epsilon}=\epsilon\alpha(2-\epsilon)(1-\alpha)$. Although a rigorous proof is lacking, we believe that this is the searched-for solution for the maximization problem. In particular, it gives $\Delta_{\cal B}^{\epsilon\to 0}(\rho^{\alpha 1})=0$ and $\Delta_{\cal B}^{\epsilon\to 1}(\rho^{\alpha 1})=E(\rho^{\alpha 1})=-\alpha\ln{\alpha}-(1-\alpha)\ln{(1-\alpha)}$, as expected. Also, we found that $\Delta_{\cal B}^{\epsilon}(\rho^{\alpha 1})=\Delta_{\cal A}^{\epsilon}(\rho^{\alpha 1})$. The behavior of the function \eqref{DeBpure} is illustrated in Fig. \ref{fig1}(b) for two values of the monitoring intensity.

\subsection{Realism-based nonlocality suppression via bilocal monitoring}
\label{swrbn}

A natural extension of the study conducted in the previous section can be proposed in terms of the quantity
\be \label{Dee}
\Delta^{\epsilon\epsilon'}(\rho):=\max_{A,B}\delta_{AB}^{\epsilon\epsilon'}(\rho),
\ee 
with $\delta_{AB}^{\epsilon\epsilon'}(\rho)$ given by Eq. \eqref{delta} and $\{\epsilon,\epsilon'\}\in (0,1)$. Notice that this quantity respects the symmetry $A\rightleftarrows B$ for $\epsilon=\epsilon'$, reduces to $N(\rho)$ as $\epsilon=\epsilon'\to 1$, and to $\Delta^{\epsilon}_{\cal B}(\rho)$ as $\epsilon\to 1$. In addition, $\Delta^{\epsilon 0}(\rho)=\Delta^{0 \epsilon}(\rho)=0$ for all $\rho$. It is instructive to observe, via Eq. \eqref{DR}, that 
\be 
\Delta^{\epsilon\epsilon'}(\rho)=\max_{AB}\left[\Delta\mathfrak{R}_A^{\epsilon}(\rho)-\Delta\mathfrak{R}_A^{\epsilon}(M_B^{\epsilon'}(\rho)) \right];
\ee 
that is, $\Delta^{\epsilon\epsilon'}$ can be viewed as a quantifier of the maximum variation in the reality change in site $\cal{A}$ upon remote monitorings in $\cal{B}$. Hence, it indeed captures aspects of realism-based nonlocality, although it is not identical to $N(\rho)$. On the other hand, since $\Delta^{1\epsilon'}=\Delta^{\epsilon'}_{\cal B}$, it is clear that, in this limit, it quantifies the realism-based nonlocality destroyed via local monitorings in this. It is, therefore, reasonable to interpret $\Delta^{\epsilon\epsilon'}$ as the amount of realism-based nonlocality that is suppressed due to monitorings conducted in each one of the distant sites.

Now, the identities \eqref{difdelta}, Theorem 1, and the result \eqref{bounds} immediately imply that
\be \label{result4}
0\leqslant \Delta^{\epsilon\epsilon'}(\rho)\leqslant \Delta_{\cal{A(B)}}^{\epsilon(\epsilon')}(\rho)\leqslant N(\rho),
\ee 
for any $\{\epsilon,\epsilon'\}\in(0,1)$ and $\rho$ on $\cal{H_A\otimes H_B}$. Equalities apply for $\rho=\rho_{\cal A}\otimes\rho_{\cal B}$. Notice the nontrivial result $\Delta^{\epsilon\epsilon'}(\rho)\leqslant \Delta_{\cal{A(B)}}^{\epsilon(\epsilon')}(\rho)$, which means that the realism-based nonlocality suppression is smaller for bilocal monitorings.

Hereafter, we restrict our analysis to the symmetrical case, $\epsilon=\epsilon'$. With a procedure similar to the one employed to derive the bounds \eqref{result1}, we use the relations \eqref{DS<} and \eqref{DS>} to obtain 
\begin{subequations}\label{result5}
\beq 
\Delta^{\epsilon\epsilon}(\rho)&\leqslant& \Gamma_{\bar{B}}^{\epsilon}(\rho)-\epsilon\,\mathfrak{I}(\bar{B}|M_{\bar{A}}^{\epsilon}(\rho))\equiv \mathrm{ub_1}, \\ \Delta^{\epsilon\epsilon}(\rho)&\geqslant&\epsilon\,\mathfrak{I}(\bar{B}|\rho)-\Gamma_{\bar{B}}^{\epsilon}(M_{\bar{A}}^{\epsilon}(\rho)) \equiv \mathrm{lb}_1,
\eeq 
\end{subequations}
where $\{\bar{A},\bar{B}\}$ are the observables that maximize $\delta_{AB}^{\epsilon\epsilon}$ in Eq. \eqref{Dee}. Once again we can conclude, from the bounds \eqref{result4} and \eqref{result5}, that while there will always be some suppression of realism-based nonlocality due to bilocal monitoring, thus precluding any increase of realism-based nonlocality under local disturbance, the total realism-based nonlocality available will never be fully destroyed, which implies no sudden death. 

\subsubsection{Example}
\begin{figure}[htb]
\includegraphics[width=\columnwidth]{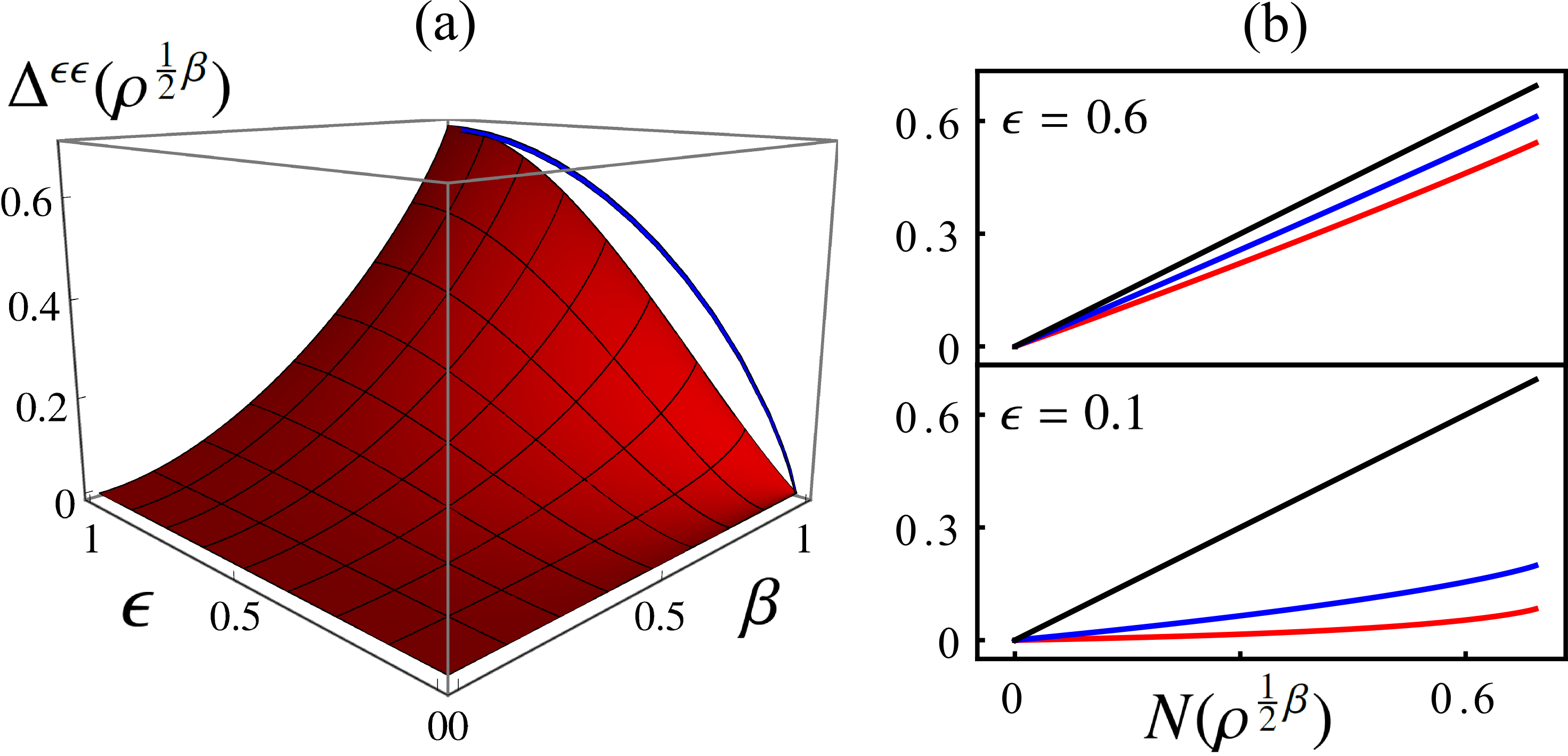}
\caption{(a) Bilocal suppression of realism-based nonlocality, $\Delta^{\epsilon\epsilon}(\rho^{\text{\tiny $\tfrac{1}{2}$}\beta})$ (red surface), as a function of $\epsilon$ and $\beta$. The blue solid line, which corresponds to $\Delta^{\epsilon}_{\cal B}(\rho^{\text{\tiny $\tfrac{1}{2}$}\beta})$, illustrates for $\beta=1$ that $\Delta^{\epsilon\epsilon}(\rho^{\text{\tiny $\tfrac{1}{2}$}\beta})\leqslant \Delta_{\cal B}^{\epsilon}(\rho^{\text{\tiny $\tfrac{1}{2}$}\beta})$. (b) Another illustration of the result \eqref{result4}. Parametric plots of $N(\rho^{\text{\tiny $\tfrac{1}{2}$}\beta})$ (top black line), $\Delta^{\epsilon}_{\cal B}(\rho^{\text{\tiny $\tfrac{1}{2}$}\beta})$ (middle blue line), and $\Delta^{\epsilon\epsilon}(\rho^{\text{\tiny $\tfrac{1}{2}$}\beta})$ (bottom red line) against $N(\rho^{\text{\tiny $\tfrac{1}{2}$}\beta})$ as a function of the parameter $\beta$, for $\epsilon=0.1$ (lower panel) and $\epsilon=0.6$ (upper panel). Local and bilocal monitorings imply suppression of realism-based nonlocality.}
\label{fig2} 
\end{figure}

We consider once again the state $\rho^{\text{\tiny $\tfrac{1}{2}$}\beta}$, given by the formula \eqref{rhoab}, and assume that the bilocal suppression is given by 
\be 
\Delta^{\epsilon\epsilon}(\rho^{\text{\tiny $\tfrac{1}{2}$}\beta})=\eta_{\sigma_r\sigma_s}^{\epsilon\epsilon}(\rho^{\text{\tiny $\tfrac{1}{2}$}\beta}), \qquad\quad \text{with}\,\, r=s\in\{x,y,z\}
\ee 
and Pauli matrices $\sigma_{r(s)}$. This assumption is based on several analytical incursions through which we have comparatively verified, for example, that smaller values occur for $r\neq s$. Also, the analytical result obtained from the conjecture above correctly leads to the limit $\Delta^{\epsilon\epsilon}(\rho^{\text{\tiny $\tfrac{1}{2}$}\beta})\to N(\rho^{\text{\tiny $\tfrac{1}{2}$}\beta})$ as $\epsilon\to 1$ and to $\Delta^{\epsilon\epsilon}(\rho^{\text{\tiny $\tfrac{1}{2}$}\beta})\to 0$ as $\epsilon\to 0$. These results can be appreciated in Fig. \ref{fig2}(a), where the smoothness of $\Delta^{\epsilon\epsilon}(\rho^{\text{\tiny $\tfrac{1}{2}$}\beta})$ with respect to monitoring strength $\epsilon$ is illustrated. In Fig. \ref{fig2}(b), full agreement with the result \eqref{result4} can be observed in the parametric plots of $N(\rho^{\text{\tiny $\tfrac{1}{2}$}\beta})$, $\Delta_{\cal B}^{\epsilon}(\rho^{\text{\tiny $\tfrac{1}{2}$}\beta})$, and $\Delta^{\epsilon\epsilon}(\rho^{\text{\tiny $\tfrac{1}{2}$}\beta})$, against $N(\rho^{\text{\tiny $\tfrac{1}{2}$}\beta})$, as a function of $\beta$ and $\epsilon$. 

So far we have discussed the behavior of realism-based nonlocality, as measured by $N(\rho)$, upon (bi)local monitoring. Our findings demonstrate that we can never destroy all the realism-based nonlocality encoded in a given preparation via this kind of weak disturbance. However, it is trivial to show that a revealed measurement of an observable $A'$ on $\cal{H_A}$ is sufficient to remove all realism-based nonlocality. This is so because upon the reduction $\rho\mapsto A'_a\otimes \rho_{\cal{B}|a}$ one obtains a fully uncorrelated state, which gives $\eta_{AB}(A'_a\otimes \rho_{\cal{B}|a})=0$ and, therefore, $N(A'_a\otimes \rho_{\cal{B}|a})=0$.

\section{Hierarchy}

The demonstrated resilience of realism-based nonlocality to (bi)local monitoring raises the question of how this quantumness notion compares to other measures of quantum correlations, for instance, quantum discord \cite{ollivier01,henderson01}, whose robustness under local disturbance, in particular to noisy channels, is well known \cite{celeri11,costa16}. In Ref. \cite{gomes18} we have already started the discussion around this topic by pointing out minimum conditions under which we can guarantee the existence of a context $\{A,B,\rho \}$ for which the context realism-based nonlocality $\eta_{AB}(\rho)$ vanishes. Now we show that realism-based nonlocality (RBN), as measured by $N(\rho)$, turns out to be the most ubiquitous type of quantumness within the set of measures involving Bell nonlocality (BN), EPR steering ($\cal{S}$) \cite{wiseman07}, entanglement (E), discord (D), and symmetrical discord (SD) \cite{rulli11}. (All these acronyms will be employed in this section for the sake of convenience.)

To construct the argument, we follow the rationale elaborated in Refs. \cite{gomes18,wiseman07}, which consists of looking for the conditions that guarantee the nonexistence of a given quantumness notion. Let us start with Bell nonlocality. As is well known, if the experimentally accessible joint probability $p(a,b|A,B)$ of finding $a$ and $b$ for $A$ and $B$ respectively measured in remote sites can be written as 
\be \label{lc}
p(a,b|A,B)=\sum_{\lambda}p_{\lambda}p(a|A,\lambda)p(b|B,\lambda),
\ee 
for whatever marginal probabilities $p(a|A,\lambda)$ and $p(b|B,\lambda)$ and hidden variables $\lambda$, then {\em local causality} \cite{bell64} is validated and Bell nonlocality is falsified ($\text{BN}=0$). If, in addition to the hypothesis \eqref{lc}, the measured probability distribution $p(a,b|A,B)$ accepts a description with $p(a|A,\lambda)=\text{Tr}(A_a\rho_{\lambda}^{\cal A})$, which refers to a model such that an observable $A=\sum_aaA_a$ is measured in site $\cal{A}$ given a local hidden state $\rho_{\lambda}^{\cal A}$, then EPR steering is falsified ($\cal{S}=0$). The hypothesis \eqref{lc} alone allows us to search for a solution for $p(a,b|A,B)$ in the entire spectrum of local hidden-variable theories and, therefore, has great power to rule out Bell nonlocality. This makes the set of Bell-nonlocal states be rather restrictive. By adding the hypothesis of local hidden states, the falsification of quantumness becomes less effective and, as a consequence, the set of steerable states becomes more inclusive. It follows that the set of Bell-nonlocal states (hereafter denoted as $\mathfrak{s}_{\text{BN}}$) form a strict subset of steerable states (denoted as $\mathfrak{s}_{\cal{S}}$), that is, $\mathfrak{s}_{\text{BN}}\subsetneq \mathfrak{s}_{\cal{S}}$. Now, let us add another restriction, namely, $p(b|B,\lambda)=\text{Tr}(B_b\rho_{\lambda}^{\cal B})$. In this scenario, where both marginal probability distributions are consistent with quantum theory, the hypothesis \eqref{lc} reduces to $p(a,b|A,B)=\text{Tr}(A_a\otimes B_b\,\rho_s)$, with $\rho_s=\sum_{\lambda}p_{\lambda}\rho_{\lambda}^{\cal A}\otimes\rho_{\lambda}^{\cal B}$ being a manifestly separable state. Clearly, the existence of such a solution falsifies entanglement ($\text{E}=0$). Again, because we have a more restrictive local model to rule out quantumness, the resulting set of entangled states, $\mathfrak{s}_{\text{E}}$, is more inclusive than the previous ones. We then obtain the relation $\mathfrak{s}_{\text{BN}}\subsetneq \mathfrak{s}_{\cal{S}}\subsetneq \mathfrak{s}_{\text{E}}$ proposed in Ref. \cite{wiseman07}.

We can add a further restriction, namely, $\rho_{\lambda}^{\cal A}=A'_{\lambda}$, where $A'_{\lambda}$ is a projector acting on $\cal{H_A}$. In this case, the local model consists of a classical-quantum state $\rho_{\mathrm{cq}}=\sum_{\lambda}p_{\lambda}A'_{\lambda}\otimes\rho_{\lambda}^{\cal B}$, which falsifies one-way discord ($\text{D}=0$). Following the argument given above, we directly conclude that the set of one-way discordant states, $\mathfrak{s}_{\text{D}}$, forms a strict superset of $\mathfrak{s}_{\text{E}}$. With the restriction $\rho_{\lambda}^{\cal B}=B'_{\lambda}$, for a given projector $B'_{\lambda}$ on $\cal{H_B}$, we then find a classical-classical state $\rho_{\mathrm{cc}}=\sum_{\lambda}p_{\lambda}A'_{\lambda}\otimes B'_{\lambda}$, for which no symmetrical discord can be found ($\text{SD}=0$). Therefore, the set $\mathfrak{s}_{\text{SD}}$ of symmetrically discordant states forms a strict superset of discordant states. We then have $\mathfrak{s}_{E} \subsetneq \mathfrak{s}_{\text{D}} \subsetneq \mathfrak{s}_{\text{SD}}$, as has been established in Ref. \cite{gomes18}. 

We are now ready to position realism-based nonlocality within the above hierarchy of quantumness notions. Let us focus on the classical-classical state $\rho_{\mathrm{cc}}=\sum_{\lambda}p_{\lambda}A'_{\lambda}\otimes B'_{\lambda}$, for which none of the aforementioned quantumness exists. By direct application of the joint-entropy theorem we can show that 
\beq \label{etarhocc}
\eta_{AB}(\rho_{\mathrm{cc}})&=&H(\{p_{\lambda}\})+\sum_{\lambda}S\left(\Phi_A(A'_{\lambda})\otimes\Phi_B(B'_{\lambda}) \right)\nonumber \\ &-&S\left(\sum_{\lambda}p_{\lambda}\Phi_A(A'_{\lambda})\otimes\Phi_B(B'_{\lambda}) \right),
\eeq  
where $H(\{p_{\lambda}\})$ is the Shannon entropy of the probability distribution $p_{\lambda}$. In order to prove that $N(\rho_{\mathrm{cc}})>0$, it is sufficient to exhibit at least one pair $\{A,B\}$ for which $\eta_{AB}(\rho_{\mathrm{cc}})$ does not vanish. This will be the case for observables $A$ and $B$ that are maximally incompatible with $A'=\sum_{\lambda}a'_{\lambda}A'_{\lambda}$ and $B'=\sum_{\lambda} b'_{\lambda}B'_{\lambda}$, respectively. In this case, $\Phi_A(A'_{\lambda})=\mathbbm{1}/d_{\cal{A}}$ and $\Phi_B(B'_{\lambda})=\mathbbm{1}/d_{\cal{B}}$, where $d_{\cal A(B)}=\dim \cal{H_{A(B)}}$. With these results, Eq. \eqref{etarhocc} readily reduces to $\eta_{AB}(\rho_{\mathrm{cc}})=H(\{p_{\lambda}\})$, which is positive for a generic distribution $p_{\lambda}$. This completes the proof that realism-based nonlocality can exist even when all the other quantumness quantifiers vanish. On the other hand, it is worth noticing that, for pure states, realism-based nonlocality becomes identical to entanglement \cite{gomes18}, just like discord does, which is a clear evidence that entanglement implies discord, which implies realism-based nonlocality. All this together indicates that the set $\mathfrak{s}_{\text{RBN}}$ of states with realism-based nonlocality ($N>0$) is a strict superset of all the other sets discussed so far. We then find the following hierarchy of quantumness notions:
\be \label{hierarchy}
\mathfrak{s}_{\text{BN}}\subsetneq \mathfrak{s}_{\cal{S}}\subsetneq \mathfrak{s}_{\text{E}}\subsetneq \mathfrak{s}_{\text{D}}\subsetneq \mathfrak{s}_{\text{SD}}\subsetneq \mathfrak{s}_{\text{RBN}}.
\ee 

\begin{figure}[htb]
\centerline{\includegraphics[scale=0.13]{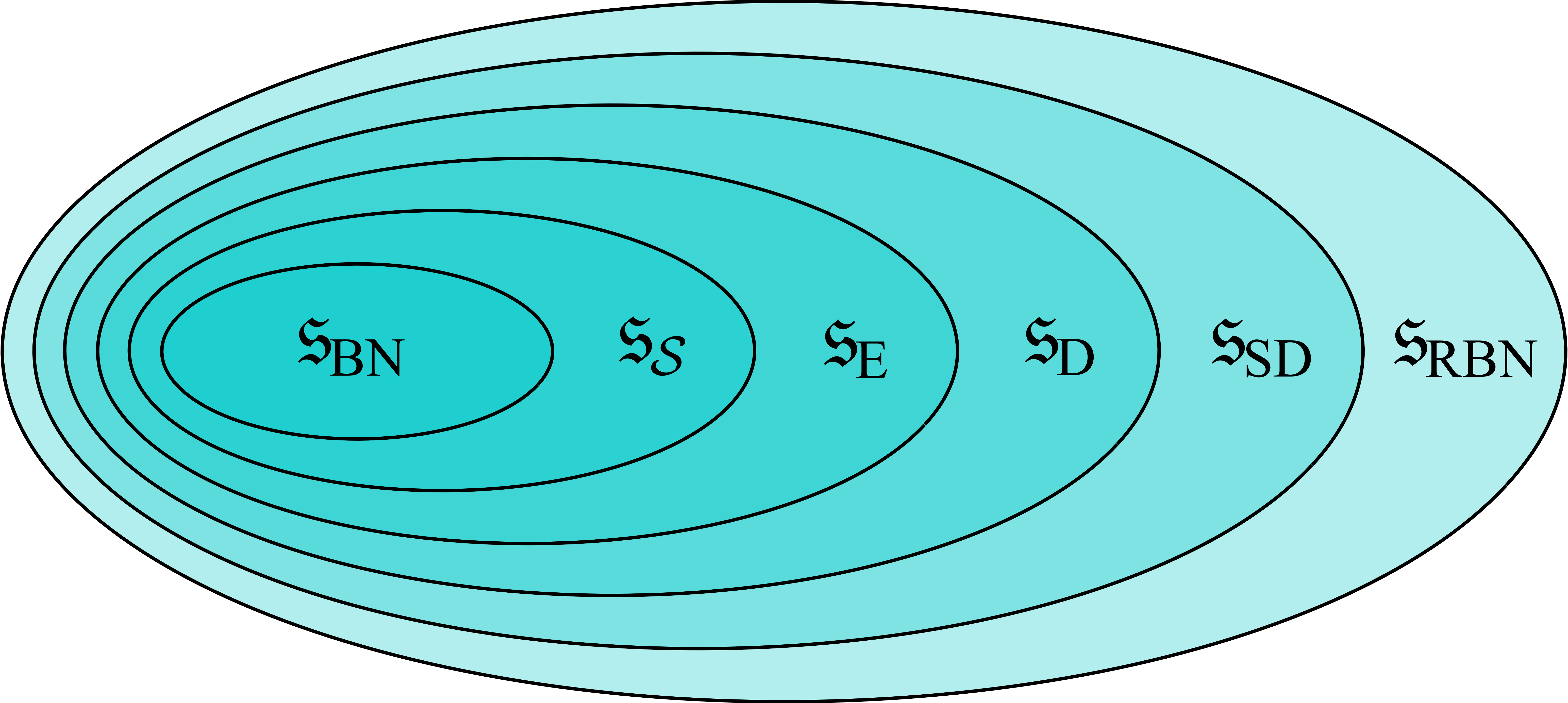}}
\caption{Representation of the hierarchy of quantumness notions \eqref{hierarchy}. $\mathfrak{s}_{\text{Q}}$ denotes the set of states possessing the quantumness Q, where Q can assume BN (Bell nonlocality), $\cal{S}$ (EPR steering), E (entanglement), D (discord), SD (symmetrical discord), and RBN (realism-based nonlocality). In this sense, realism-based nonlocality is the most ubiquitous type of quantumness, being absent only for fully uncorrelated states.}
\label{fig3} 
\end{figure}

This result attests that realism-based nonlocality is the most ubiquitous species of quantumness within the class here considered (see illustration in Fig.~\ref{fig3}). Indeed, according to the theoretical evidences collected so far, we can safely state that it vanishes only for fully uncorrelated states. This is so because, being based on the notion of irreality, realism-based nonlocality is highly sensible to the incompatibility of observables, thus existing even for symmetrically nondiscordant states. 

\section{Summary}
In this work, we investigated further properties of the recently introduced realism-based nonlocality \cite{gomes18}---a notion of nonlocality that emerges from the violation of the hypothesis that a remote disturbance in site $\cal{B}$ does not change the degree of irreality in site $\cal{A}$. Specifically, we studied the resilience of this quantumness notion to (bi)local weak measurements, which were implemented through the monitoring map \cite{dieguez18}. Since monitoring is a particular class of local operations, our results show that realism-based nonlocality does not increase under these local actions, which is a desirable feature of quantum-correlation quantifiers. In particular, we demonstrated that, apart from the regime of strictly projective measurements, realism-based nonlocality can never be fully destroyed. Therefore, unlike Bell nonlocality, EPR steering, and entanglement, realism-based nonlocality is not susceptible to sudden death. As follows from the hierarchy demonstrated in Sec. IV, this behavior is consistent with the fact that no sudden death has ever been observed for discord or symmetrical discord. The remarkable robustness of realism-based nonlocality naturally raises questions about eventual applications of such nonlocal aspects in physical tasks, such as those of information theory and quantum thermodynamics. These questions are, however, left open for future work. 

\section*{Acknowledgments}

This study was financed in part by the Coordena\c{c}\~{a}o de Aperfei\c{c}oamento de Pessoal de N\'{\i}vel Superior--Brasil (CAPES)-Finance Code 001. R.M.A. acknowledges partial support from the National Institute for Science and Technology of Quantum Information (INCT-IQ/CNPq, Brazil).


\end{document}